# Implications for Improving Accessibility to E-Commerce Websites in Developing Countries - A Study of Hotel Websites

Arunasalam Sambhanthan
School of Computing, University of Portsmouth, United Kingdom
arunsambhanthan@gmail.com

Alice Good
School of Computing, University of Portsmouth, United Kingdom
alice.good@port.ac.uk

**Abstract**

This research explores the accessibility issues with regard to the e-commerce websites in developing countries, through a study of Sri Lankan hotel websites. A web survey and a web content analysis were conducted as the methods to elicit data on web accessibility. Factors preventing accessibility were hypothesized as an initial experiment. Affecting design elements are identified through web content analysis, the results of which are utilized to develop specific implications for improving web accessibility. The hypothesis tests show that there is no significant correlation between accessibility and geographical or economic factors. However, physical impairments of users have a considerable influence on the accessibility of web page user interface if it has been designed without full consideration of the needs of all users. Especially, visual and mobility impaired users experience poor accessibility. Poor readability and less navigable page designs are two observable issues, which pose threats to accessibility. The lack of conformance to W3C accessibility guidelines and the poor design process are the specific shortcomings which reduce the overall accessibility. Guidelines aim to improve the accessibility of sites with a strategic focus. Further enhancements are suggested with adherence to principles, user – centered design and developing customizable web portals compatible for connections with differing speeds. Re-ordering search results has been suggested as one of the finest step towards making the web content accessible for users with differing needs. A need for developing new design models for differencing user groups and implementing web accessibility strategy are emphasized as vital steps towards effective information dissemination via e-commerce websites in the developing countries.

## 1. Introduction

Accessibility is a critical factor for e-commerce success. Particularly, the accessibility of e-commerce websites for users with impairments is an evolving research area in the contemporary academia. Apart from this, the accessibility of e-commerce websites in developing countries could be influenced by several other factors such as economic background and geographical location of users. Therefore, it is a vital requirement for web concept designers in the developing countries to consider the above factors in order to ensure the successful dissemination of business information through e-commerce sites. However, the means of designing accessible web interfaces is still remaining as an open question with regard to developing countries. By not adhering to accessibility guidelines, e-commerce businesses are inadvertently excluding a potentially large demographic. According to a report released by United States Census Bureau on year 2005, 6.8 percent of the population age 15 years and older records sight and hearing impairments. This amount increases to 21.3 percent when considering the population above age 65 years. 8.2 percent of the same population records having difficulty in grasping objects – which affect the use of mouse (Americans with Disabilities, 2005). This clearly shows the significance and gravity of this research. According to Australian Bureau of Statistics the 35.9 percent of the Australian population is subjected to hearing impairments in year 1993. The United Nations Statistics Division reports that most of the population of above 40 years age has subjected to some form of disabilities. This shows the emerging need for considering disabled population during e-commerce website design process.

This paper aims to formulate implications for improving accessibility to e-commerce websites in developing countries through a study of Sri Lankan hotel websites. Specifically, formulating implications for web designers would be a vital contribution to the body of knowledge. Also, the restricting factors on reaching potential customer base could be effectively tackled through implementing these implications. The specificity of this research has been focused to Sri Lankan hotel websites considering the contextual relevance. Arguably, Sri Lankan hotel websites have a greater need for accessibility due to the reservations made by tourists from different physical, economic and geographic backgrounds. The current economic boom, opened by the post war scenario of the island further signifies the emerging need for designing accessible e-commerce sites, for effective web based promotion.

## 2. Literature Review

Webster's dictionary defines accessibility as "the quality of being accessible, or of admitting approach." (Webster's Dictionary - Revised Unabridged Dictionary, 1996, 1998 MIRCA, Inc) This definition serves as an introductory note for the term, but insufficient to explicit the innate meaning of accessibility in its practical form. On the contrary, W3C providing a more contextual definition for the term, states "Web Accessibility is a term used to identify the extent to which information on Web pages can be successfully accessed by persons with disabilities including the aging" (W3C, 2000). However, this definition only includes the physical disabilities, but fails to include some other important dimensions of this multi-faceted word. Particularly, accessibility could means a lot in the context of developing countries which are far behind developed countries in terms of economic condition. Consequently, Good (2008, p. 16) defines accessibility in a more versatile manner which definition encompasses the contextual requirements of developing countries. The definition goes as, "A website is said to be accessible when anyone, regardless of economic, geographic or physical circumstances, is able to access it". This means, the ease in which, people from different economic backgrounds, people living in different geographic zones and people having different physical impairments could access the web'. Hence, for this particular study, accessibility could be defined as '*the ease in which people with disabilities, people from different geographic regions and people having different internet connections could access the websites.*'

### 2.1. *Economic Factors*

The economical aspect of accessibility could be looked in two ways, (1) the category of internet connection - which depends on the economic background and usage frequency of customer, and (2) the time spent on information search which mostly depends on the time availability of consumer. Both these factors could have an interrelationship and will have a direct effect on accessibility. However, a predisposition has been taken as the economic background could be directly measured through the type of connections people have. Arguably, the people with high economic standards would choose Asymmetric Digital Subscriber Line (ADSL) or broad band connections while the people with low income chooses dial up connections, in the context of a developing country.

The type of internet connection determines the speed of browsing, which influences the time a consumer spends on browsing. The less speedy the connection, the more the consumer will get discouraged and vice-versa. On the other hand consumer's average browsing time per day determines the category of connection they hold in the context of a developing country. However, search engine optimization - an evolving area in tourism research deals with this issue to a certain extent. The degree of filtration and the sequence of listings done by search engines will influence the accessibility of tourism sites. Law & Haung (2006) indicates that 47.4% users only check the first three pages of the search results listing. The above finding has not specified the number of results listed per page. However a recent study of Xiang *et al* (2008) moderates the previous finding, stating that travelers are only retrieving a tiny fraction of the huge list of search results. Evidently, search result listing is a dominant factor which could influence the decision behavior of consumers. But, it could be argued whether the curious consumer will not spend more time to get the best out of his search. Also, the above rule might be inapplicable for frequent customers as they would directly go to the hotel web page to check the information. However the information search behavior of consumers will trigger to have a look for 'innovation' in each hotel's website. On the other hand, hotels will not be able to survive in today's knowledge economy with their 'obsolete' ways of promoting products. Consequently, being in the forefront of search results listing and using a customer centric approach in wording websites are vital for getting the customers to stay with the site.

## 2.2. Geographic Factors

The geographic aspect is another major factor which determines accessibility. Accessing the website from rural areas with poor bandwidth is another challenge. Utilization of mobile commerce is another evolving area in tourism to solve this issue. According to Hyun *et al* (2009, p. 158) it is critical for hotels to utilize mobile commerce in tourism promotion to survive in an era of increasing mobile adaptation. In addition Buhalis (2008, p. 612) claims mobile technology increases the accessibility of digitally excluded communities. However, investing on mobile technology could pose a number of challenges to hotels. Pizam (2009, p. 301) indicates that the current global financial crisis has adversely affected the hospitality industry and tends the income of hotels to be reduced over time. On the other hand, Buick (2003, p. 244) identified customer demand as a pulling factor in IS/IT investment. Will it be advisable to invest on mobile technology in the light of the above contradicting dilemma? But the competitiveness gained through new technology could be powerful than any of the above factors. Supportively Kozak *et al* (2005, p. 7) ascertain that the ability to gain IT access is the deciding factor of first and second class tourist destinations, organizations and customer groups.

## 2.3. Physical Factors

Although there are differing definitions and literary interpretations from several perspectives available for the term disability (W3C, 2001, HMSO, 1996), Nielsen (2000) interprets the term disability as 'any difficulty experienced in interacting with a computer system.' The interpretation matches very well with the scope of this research which focuses on the accessibility of e-commerce websites. Therefore the term disability could be well aligned with the research theme by narrowing down Nielsen's interpretation of disability to the context of web based systems. Hence disability could be defined as '*any difficulty experienced in interacting with an e-commerce website during the course of pre-purchase, consumption and post-consumption stages of an e-commerce business.*' Visual and mobility impairments are the main types which influence the accessibility of websites. Hence, both these impairment types and the effect of those impairments on web accessibility are discussed below.

## 2.3.1. Visual Impairments & Accessibility

According to Nielsen (1996) visual impairments cause most serious accessibility issues. Visual impairments are of many types such as complete blindness, partially sighted, etc. Regardless of the impairment type, the total population of people with visual impairments is predicted to be increasing rapidly in a rate of 2 million per year worldwide (WHO, 2002). When determining how disability affects accessibility, blind people are considered to be unable to "see", or rather access a web page, if they are not provided with assistance in the form of a screen reader or an audible web client. Although these aids are useful, their application is somewhat limited. The use of graphics without necessary guides or aids results in a page being categorically unreadable. Given that a vast majority of web sites are designed to be visual, nearly 50 million people are then automatically excluded (WHO 2001). The visual impairments have been categorized into two as, blind and low vision throughout the following discussion.

According to W3C, people with blindness will encounter a number of barriers in using the web which include:

- Images that do not have 'alt' text
- Complex images (e.g., graphs or charts) that are not adequately described
- Video that is not described in text or audio
- Tables that do not make sense when read serially (in a cell-by-cell or linear fashion)
- Frames that do not have "NOFRAME" alternatives or that do not have meaningful names
- Forms that cannot be tabbed through in a logical sequence or that are poorly labeled
- Browsers and authoring tools that lack keyboard support for all commands
- Browsers and authoring tools that do not use standard applications programmer interfaces for the operating system they are based in
- Non-standard document formats that may be difficult for their screen reader to interpret

*(W3C 2001, How People with disabilities Use the Web)*

The problems summarized by the W3C underline the severity of the challenge facing the blind. Their access to web-based information can be impaired in any number of ways. Oftentimes, access is simply impossible. On the other hand, the people with low vision, often known as partially impaired, experience certain barriers in browsing the web which are also audited by W3C.

- Web pages with absolute font sizes, i.e. that do not enlarge or reduce easily

- Web pages that, because of inconsistent layout, are difficult to navigate when enlarged, due to loss of surrounding context
- Web pages, or images on Web pages, that have poor contrast, and whose contrast cannot be easily changed through user override of author style sheets
- Imaged text that cannot be re-wrapped
- Also many of the barriers listed for blindness, above, depending on the type and extent of visual limitation

*(W3C, 2001, How People with disabilities Use the Web)*

Hence, all visually-impaired people experience some difficulty in struggle with accessing web-based information. While blind people may have greater difficulty than those who are partially sighted, the problems are similar. A lack of consistency in web design is a key problem along with the highly visual nature of most web pages (Good, 2008).

### *2.3.2. Mobility Impairments & Accessibility*

People with physical or motor impairments may experience some difficulty accessing the web, however, these disabilities often manifest themselves differently than visual impairments. However, it is essential to distinguish between the disabilities which cause inaccessibility to websites. Arguably, a person with wheelchair is classified as mobility impaired but not necessarily experience difficulty accessing the web. However, people with dexterity impairments, either as a result of a neurological disorder or arthritis (Emiliani, 2001), who are affected.

In general, problems consist of difficulties in manipulating the mouse or performing precise movements with a tracking pad. As well, there may be an inability to use the keyboard, in particular holding down two or more keys simultaneously. According to Nielsen (1996), ten years ago image maps were the main cause of access difficulties simply because they required such precision in usage. Since this time, page navigation has become increasingly complex with buttons and links becoming smaller and requiring a greater degree of precision and motor skills (Hackett *et al*, 2003). There are a number of other problems this group may face. Motor disabilities manifest themselves in a variety of ways. The W3C give the following examples of the problems that this category may experience in accessing web-based information.

- Time-limited response options on Web pages
- Browsers and authoring tools that do not support keyboard alternatives for mouse commands
- Forms that cannot be tabbed through in a logical order (Note: "Tabindex" solution not yet well supported in browsers.)

*(W3C 2001, How People with disabilities Use the Web)*

Hence, as is the case in almost every category, users with motor disabilities suffer from a lack of consistency in web design. As well, some features of web pages fail to allow for different response times. It is also reputed in several researches that the access to web based information is inadequate for disabled people (Hackett *et al*, 2003; Good, 2007; Parmanto & Zeng 2003; Bevan & Ahmed 2007). Evidently, the influence of physical circumstances has a major role in accessibility. Eichhorn *et al* (2008) argues more sophisticated understandings of differential needs and appropriate sources as the most crucial step in enabling disable tourists to access tourism information. Apart from the influence of aforementioned inherent barriers each users have based on the type of their impairments, there will also be an impact if web design is poor. For example users may find it difficult to use the ecommerce site due to the need of making precise mouse movements, small links, fixed fonts, poor contrast between foreground and background colors, etc. Hence, these issues have to be taken into consideration at the design phase.

## 2.4. Accessible Design

The primary reason for most of the accessibility problems is poor design, which could be complemented through user involvement and feedback. Standards for achieving content accessibility are widely available; however, despite this, there is a marked absence of compliance from the majority of web designers. There are a number of tools available to assess the accessibility of individual website, but these measures are rarely used. Hence it is inevitable to have a design strategy for improving accessibility.

### 2.4.1. What Causes Accessibility Problems?

A number of factors contribute to problems with accessibility. Initially, the general consensus was that poor design and coding errors were the primary causes, these factors contributed to the majority of problems related to accessibility (Nielsen, 1996; Shneiderman, 2000). More recently, other factors have also been identified. A lack of consideration in design for all users is a contributing factor. Early research suggests that designers fail to take into consideration the wider range of user needs (Preece *et al,* 2001; Clement *et al,* 1999; Hackett *et al,* 2005; Chiang, Cole, Gupta, Kaiser & Starren, 2005). While website accessibility has improved somewhat since that time, problems with accessibility still persist for disabled users (Leitner & Strauss, 2008) Researchers are unanimous in agreeing that a lack of user involvement and feedback during the design process is a major contributing factor in accessibility problems. It is, however time-consuming to involve end-users throughout the design stage. Over time, researchers have identified possible solutions. Early research suggests that designers should employ a wider range of scenarios in order to understand usage (Shneiderman, 2000; Preece *et al,* 2001). More recent research, which focuses on the visually impaired, suggests assistive computer technologies may provide solutions (Chiang *et al* 2005). Similarly, researchers seeking to address accessibility issues for those who are hearing impaired suggest assistive technology, specifically the IBM Viascribe which acts as captioning tool, may offer a feasible solution (Bain, Basson, Faisman & Kanevsky, 2005).

According to Lynch and Horton (2001), only part of the problem can be attributed to a lack of consideration for user needs. The real problem is that designers fail to take into account the equipment used by the majority of people. Oftentimes, designers work from high-quality equipment; however, the end-user does not always enjoy the same high standards. A difference between screen size or web browser can make a significant difference to a site's accessibility. While a site may work well on a high definition screen using a particular browser, it may not translate to the budget laptop used by the average person. There is a solution, however, 34 and the recommendation to designers is simple: "Don't design for your machine, design for your average reader" (Lynch & Horton, 2001).

There is considerable evidence that the consultancy process during website development leaves much to be desired. A significant number of designers fail to consult their intended users at any point during the design process. This frequently results in problems. While it is possible to design systems without involving users, this is not necessarily the best option. It may be cost-effective and less time consuming in the short term, however, it may leave designers vulnerable to charges of discrimination. Research suggests that the insular approach to design is markedly unsuccessful when dealing with users whose needs encompass a wider range than those envisaged by the designer, in particular users with impairments (Shneiderman, 2000). The Human Rights and Equal Opportunity Commission has outlined the main elements that have an impact on accessibility. They include:

- Lack of Alternate (ALT) text: this is a common cause of accessibility problems, where images are not supplied with alternative text
- Non-textual elements
- Tables and forms
- Image maps and navigation
- Color
- Text and Paragraph Formatting

*(The Human Rights and Equal Opportunity Commission, 2002)*

While this list identifies some of the same problems, it is not as complete as W3C. It does not take into consideration the needs of specific user groups and does not provide the same level of guidance.

### *2.4.2. Navigation and its Effect on Accessibility*

Navigation plays an important role in providing accessibility. It does, however, require careful consideration. The ease with which different users navigate a site reflects how carefully the site has been designed. A successful designer will aim to create a website that allows different kinds of users to interact and move about the site effectively and efficiently without placing undue stress on the user. While considerable focus has been placed on content, the importance of navigation cannot be ignored. According to Nielsen (2001), good web site navigation ensures a high degree of usability. Unfortunately, too often this proves to be yet another area in which designers are guilty of a certain degree of negligence. There are a number of navigational devices available, such as bars, breadcrumb trails, hypertext links, hierarchical maps, site maps and tables of contents. However, not all of these devices are useful or indeed useable. Navigation bars, which consist of a collection of hyperlinks, are a popular choice and can be implemented in a variety of ways. These include textual and graphical menus or pop-up menus. To ensure accessibility, however, those that use graphics need to comply with appropriate application of the <ALT> attribute. The history of items viewed and changes in hyperlink colors also need to be considered.

### *2.4.3. Standards and Guidelines*

When the Internet first came into being, there was little thought or consideration given to the needs of special user groups (Laux, 1996). Over a decade later, not all user needs are considered in the design process. There are an infinite number of websites that have been designed in such a way as to make access by users with special needs difficult if not impossible. In its original incarnation, the Internet used a text-based medium that would have been easily accessible by today's assistive technologies. The Internet, however, has since transformed into a multimedia environment. In doing so, it created a digital divide between general user groups and those with disabilities (Waddell 1999). Previously, people with visual disabilities were able to access the Internet using screen readers, which would audibly read the text on the web page. Today, unless programmers incorporate accessible web design, graphic based web pages create insurmountable barriers. Since the birth of the World Wide Web, a number of organizations and individuals have proposed various guidelines and standards. Only one set of guidelines, however, has been adopted by W3 (Sullivan 2000). As part of their mission to increase accessibility, the W3C introduced the Web Accessibility Initiative (WAI) in 1997. As a result, the Web Content Accessibility guidelines were developed later on in 1999 to assist designers in creating accessible content for users with disabilities.

### 2.4.3.1. Web Content Accessibility Guidelines

*Guideline 1: Provide equivalent alternatives to auditory and visual content.*
*Guideline 2: Don't rely on color alone.*
*Guideline 3: Use mark-up and style sheets and do so properly.*
*Guideline 4: Clarify natural language usage.*
*Guideline 5: Create tables that transform gracefully.*
*Guideline 6: Ensure that pages featuring new technologies transform gracefully.*
*Guideline 7: Ensure user control of time-sensitive content changes.*
*Guideline 8: Ensure direct accessibility of embedded user interfaces.*
*Guideline 9: Design for device-independence.*
*Guideline 10: Use interim accessibility solutions.*
*Guideline 11: Use W3C technologies and guidelines.*
*Guideline 12: Provide context and orientation information.*
*Guideline 13: Provide clear navigation mechanisms.*
*Guideline 14: Ensure that documents are clear and simple.*

*(Taken from WC3, Web Content Accessibility Guidelines)*

Sites that have complied with these standards receive certification of accessibility in the form of an embedded symbol. In addition, there are three levels of compliance with these guidelines, ranging from basic to optimum. There have been a number of updates to the guidelines since 1999 with the most recent in 2006.

### 2.4.4. Tools to Check Accessibility Standards

There is a range of tools available to check the accessibility of websites, and the number has grown over recent years. The W3C provides a variety of devices for designers, including evaluative and reparative tools. Also available are filter and transformer tools, which help the user, adapt a page or offer assistance on where to find a more accessible browser. A common example of this type of tool is known as Bobby (Bobby 1995). Bobby is a free service provided by the Centre for Applied Special Technology (CAST). When a Uniform Resource Locator (URL) is provided, Bobby checks the page against WAI guidelines. It then identifies potential problems and indicates the nature of the problem with a violation symbol. In addition, Bobby provides a description of the problem and suggests possible solutions. Bobby divides problems into three main categories: errors, warnings and suggestions. An error indicates that the page is inaccessible; potential problems such as missing ALT text or invalid attributes are flagged. A warning indicates a more serious problem. For example, an error that needs fixing but would affect the overall design, i.e. Hyper Text Markup Language (HTML) forms or tables. The suggestions the suggestions simply advise which components could improve accessibility if changed. Bobby has been acquired by Watchfire and is now available as a commercial application. Bobby 5.0, acquired by Watchfire in 2002, continues to provide many of the same functions first developed by the Centre for Applied Special Technology (CAST) however, the product appears to be now targeted at a business audience. The hefty price tag and increased scanning capabilities suggest that this product is meant as a diagnostic tool (Watchfire 2005). The promotional literature addresses the business user offering guidance on how to make websites more accessible. An online URL checker is still available, however, free of charge. It performs a scan of website, identifying possible problem spots.

## *2.5. Reordering Search Results as an approach to improving access to more accessible web pages*

If disabled users cannot access the ecommerce websites due to poor accessibility as a result of poor design, then a significant percentage of society will be excluded from these sites. This then has a financial implication for the businesses if their ecommerce sites are being rejected in favor of more accessible one. A recent research suggests a new method for improving accessibility to websites through re-ordering search results according to user needs. Good (2008) investigated a means to automate the rating of web pages according to their accessibility to specific user groups. These groups include visual impairments, mobility restricted and dyslexia. The research identified an integrated, user-centered study. In the requirements phase,, user defined ratings of web page elements that are known to affect accessibility for specific user groups were obtained. These ratings form the basis of a set of algorithms that aimed at analyzing the accessibility of a web page and assign a rating. Web Pages are then re-ordered according to how accessible they are to a particular user group.

This research would have further implications on the businesses because their sites would be ranked according to their accessibility, and thus may not feature on the first page of search results. Also, this research shed light on adapting a universal design of web pages, which could include a wide range of potential target audience. However, the profundity of Good's (2008) work could be criticized for its limitation to cover the whole population of disabled consumers. Also the sample selected by Good (2008) is very small and could be claimed as inadequate to generalize the findings. However, Good (2008, p. 134) defended her research claiming that excellent results could be achieved using small groups (Nielsen, 1996) Furthermore, all visually impaired users will face difficulties in reading the letters with small font size and blind users will also experience issues that are the same as other blind users.

## *2.6. Summary*

The findings from the literature review report on the issue of accessibility. In spite of legislation and guidelines, users are still facing difficulties accessing and perceiving web based content. User designed approaches to design have been advocated for many years yet research still suggests a lack of adherence to these methods. The impact upon businesses through a lack of design compliance is heavily publicized. There is even a method to simply bypass non- accessible pages by re-ordering the pages according to accessibility (Good, 2008). For hotels to succeed in a competitive online environment, careful consideration to all users should be incorporated into the design of any hotel based website.

## 3. Methodology

A user centric approach was adapted for this research. The data collected through a web survey and web content analysis. Initial hypothesis testing was conducted based on the data collected through web survey. The hazardous design features which prevent accessibility were triggered through the web content analysis.

### 3.1. Selected Methods

Traditional paper based survey is feasible for small target groups as it could be easily distributed and collected within limited time rage. Especially for a certain group of staff belong to an organization could be distributed and collected with paper based survey within few minutes. In contrast, these surveys are comparatively costly and could not be used for geographically dispersed large samples. Also, the researcher's effort and physical presence are deemed in traditional surveys. Alternatively, postal surveys could be used to collect data from geographically dispersed samples, without the physical presence of researcher. However, the cost and time will be comparatively high in postal surveys. Apart from the above challenges, vague questions will prevent effective responses and any in-mid modification cannot be done in surveys as it will affect the data analysis through inducing variability in responses. Despite the above factors, poor response rate is an identified inherent limitation of survey research.

Justifiably, web based surveys could be used to gather data from geographically dispersed large samples with less time, cost and effort. Buick (2003) and Iliachenko (2006) composed their studies with pilot tested electronic surveys. Piloting will help to eliminate any inadequacies or usability issues with the research instrument. Also Buick (2003) used incentives in the form of a prize draw and records 30% response rate, in fact monetary incentives were criticized by research community for its inappropriateness in business surveys. However, providing any form of incentives will not be feasible in this research, as the allocated budget is not sufficient for incentives. But the response rate could be increased through effectively promoting the survey. Additionally, a web survey will be easy to create, handle and dispose anytime anywhere due to the portability and flexibility it offers. Especially, the ability to send and promote web surveys through electronic means is a contextual advantage for this research in terms of time and scope concerns. Also, the survey intends to collect data on the accessibility of websites, which needs the user's intervention with the World Wide Web (WWW). Web survey is more flexible for this task compared to postal or other types of surveys as, the respondent will feel flexible of filling the survey while performing the tasks given. The above reasons were directed the choice of web survey for this research.

Observations have the potential to obtain highly scientific results as it allows scrutinizing the exact habit instead of ideal behavior. Any misrepresented data could be effectively validated through observations. Additionally, a structured observation will allow tracing a certain set of specific elements. Also the specific accessibility features could be easily observed through a web content analysis. Reasonably, the web content analysis will help to identify the shortcomings in terms of specific features, which could be directly used in developing implications. Therefore, observation method in the form of a web content analysis has been selected to validate the interview results.

*3.2. Research Design*

The survey intended to trace data on accessibility of the current sites for users from different geographic, physical and economic backgrounds. The instrument was designed to collect data such as the type of internet connection and country of residence. This was intended to measure the accessibility of site by eliciting basic information on the internet browsing pattern of users. The above data were utilized to statistically measure the impact of physical, geographic and economic factors on accessibility.

A brief introduction about research provided at the front page. Rationales for questions were commonly provided. Individual rationales were provided for sensitive questions such as special needs and email address. The special needs were classified into four main categories and a number of impairments falling under each category were described in the research website and linked with survey.

Refer sample groups section for explanation on special needs classification. This was done to get the informed consent of respondents for filling the survey. Navigational links to the sample list were placed in every page of the survey. This was to enable the user re-browsing sample site at any stage. The windows of sample sites set to be opening in new window to reduce usability problems. Jumping commands was programmed within sub sections. Especially, the respondents answered they do not have a special need were set to be automatically skip the question asking to specify the kind of special need.

Likert style questions were designed to trace the user opinion on web content criteria. Opinions ranging from strongly disagree to strongly agree were assigned with codes 1-5 respectively. The middle response was set to be equal to average. The positive and negative user experience were traced through yes or no and coded as 1and 0 respectively. Rating the opinions based on an arithmetic series was the rationale for coding.

A website was created to provide the preliminary information to survey respondents. The web based survey was piloted with 7 participants representing all three sample groups to improve the usability of it. The improvement of language, inclusion of navigational links and the incorporation of rationale for questions were the notable outcomes of pilot study. The hotels falling under western region of Sri Lanka (22) were selected and the hotels with e-commerce websites were shortlisted based on a consequent Google search.

The number of questionnaires to be promoted was processed as 120. In fact the maximum number of dependent variables was determined as 4. Altogether 80 responses were estimated deciding 20 responses per variable. To reach 80 responses it was decided to gather 120 responses considering 40 invalid responses. Considering the above outcome the survey was planned for one full week. Links of all 22 hotel sites were given to the respondents and they were asked to browse one website on their own choice and to record their experience.

The elements rated by users analyzed through a quantitative approach. The analysis of statistical variance was undertaken to check the validity of each sub hypotheses. Analysis of Variance (ANOVA) test of user ratings was undertaken using online analysis software. The web content of selected five hotels was analyzed against a checklist prepared based on a randomly selected criterion from W3C web accessibility guidelines. Apart from this an accessibility analysis was conducted for the same sites using an automatic accessibility evaluation tool. This approach is to ensure the accuracy of results. Arguably, W3C states that the automatic assessment needs to be validated with manual ones due to the unreliable nature of automatic tools. Apart from this, short interviews were conducted with hotel managers to explore the organizational initiatives with regard to improving accessibility.

### *3.3. Sample Groups*

The study was conducted in the natural environment of user. 120 participants from 14 countries participated in the survey. The sample was classified into different groups check the accessibility of hotel websites for users from varied economic, physical and geographic backgrounds. User samples included connection types [ADSL broadband and any other], impairments categories [visual and mobility impairments] and geographic regions [Sri Lanka and foreign users].

The connection types were classified into three main categories namely ADSL, broadband and any other connections. Any other connection category includes dial up connection as well. The first two connections represent people having good bandwidth and speedy internet accesses, while the other category denoted people with below average bandwidth.

The geographic destinations have divided into two categories namely, Sri Lankan and foreign users. Rationale for which was to assess the accessibility for both within and outside Sri Lanka. Justifiably, the research not intends to assess the country wise accessibility of sites, instead accessibility for different geographic destinations.

Special user needs have classified into four main categories namely, visual, auditory, cognitive/language and mobility. The categorization of visual and auditory needs includes all the impairments related to vision and hearing. The cognitive/language needs include impairments related to language, speech and cognition. The mobility includes dexterity and motor impairments. Good (2008, pp. 28) classified impairments into the above categories to demonstrate the relationship between design features and accessibility. However, visual and mobility impairments are considered under this study as both these impairments have a significant influence over the accessibility of sites.

The above classification was adopted as it gives a comprehensive method for classifying disabilities with regard to internet accessibility. Arguably, one of the objectives of this study is to measure the accessibility of sites for users with different impairments, but not to measure the internet accessibility for individual impairments. However, the auditory impairment was eliminated from the list in later stages as there were not enough responses recorded from participants representing that group. A brief description of the above categorization was given in the research website and linked with survey to enable respondents understand the method followed for classification.

While there may be some criticism on selecting a sample of generic users instead of tourists, it is defendable in this instance. The purpose of study is to measure the accessibility of sites. Accessibility of a site is largely depending on aspects such as users' geographic, economic and physical backgrounds and site design (Good 2008). In fact, the accessibility of a site would be same for a tourist and non tourist user.

## 4. Results

Table 1 presents the accessibility ratings of websites provided by users. Participants asked to rate the accessibility of sites. The correlation between accessibility and living zone, type of connection and physical impairments were measured accordingly. Notably, different living zones represent geographic diversity while different types of connections and different impairment types represent economic and physical diversity of samples. The, user ratings were aggregated and presented as a table. The likert style assessment was adapted and responses were coded as follows. (SD: Strongly Disagree, D: Disagree, NAND: Neither Agree nor Disagree, A: Agree, SA: Strongly Agree)

| Classification | SD | D | NAND | A | SA |
|---|---|---|---|---|---|
| Sri Lanka | 1 | 1 | 9 | 20 | 3 |
| Other Countries | 0 | 1 | 22 | 19 | 4 |
| Mobility | 0 | 1 | 1 | 2 | 0 |
| Visual / Sight | 0 | 1 | 0 | 2 | 1 |
| ADSL | 0 | 0 | 15 | 18 | 4 |
| Broad Band | 1 | 2 | 15 | 19 | 3 |
| Dial Up | 0 | 0 | 1 | 1 | 0 |
| Any Other | 0 | 0 | 0 | 1 | 0 |

Table 1: Accessibility Ratings of Users

Country wise accessibility scores show most of the responses are above 'neither agree nor disagree' category. There are no users with auditory impairments. Most of the users with cognitive / language and mobility impairments have rated above average score for accessibility. However a visually impaired user disagrees with the accessibility. Most of the responses for connection wise accessibility show above average category. Thus, generally the websites could be said to be accessible for users from varied geographic, economic and physical backgrounds, except the tiny fraction of negativity recorded from users with visual and mobility impairments. However, the statistical mean and median for each rating were calculated for further analysis.

## 5. Statistical Analysis

The following section presents and discusses the statistical results conducted for connection wise, country wise and impairment wise accessibility ratings of users.

### 5.1. Connection wise Accessibility

| Source of variation | Sum of squares | d.f. | Mean squares | F |
|---|---|---|---|---|
| Between | 0.3574 | 2 | 0.1787 | 0.391 |
| Error | 30.63 | 67 | 0.4571 | |
| Total | 30.99 | 69 | | |

Table 2: Statistical Analysis of Connection wise Accessibility Scores

According to the information presented in table 2, the probability of this result assuming null hypothesis is 0.68. Therefore the null hypothesis is valid and the research hypothesis cannot be accepted. Hence, the type of connection does not have any significant influence over the accessibility of sample websites. Therefore, the economic background of the user does not need to be a deciding factor of accessibility according to the above hypotheses.

### *5.2. Country wise Accessibility*

| Region | Mean | Standard Deviation | Median |
|---|---|---|---|
| Sri Lanka | 3.66 | 0.614 | 4 |
| Foreign Countries | 3.54 | 0.711 | 3 |

**Table 3: Statistical Analysis of Country wise Accessibility Scores**

In table 3, the regions have been divided into two main groups as Sri Lanka and foreign countries, and a hypothesis was tested to scrutinize whether the geographic diversity has any significant effect on the accessibility of websites. The results indicate that the probability of these results assuming the null hypothesis is 0.47, which is lesser but much closer to the middle point. However, the mean and median of the accessibility scores shows a more than average score for accessibility from differing regions. Especially, the raw data reveals that almost 90% users rated 'above average' for accessibility. Hence, it is concluded that the null hypothesis is valid and there is no significant influence on accessibility by the regional diversity.

### *5.3. Accessibility for Users with Special Needs*

| Source of variation | Sum of squares | d.f. | Mean squares | F |
|---|---|---|---|---|
| Between | 0.5 | 2 | 0.25 | 0.2647 |
| Error | 8.5 | 9 | 0.9444 | |
| Total | 9 | 11 | | |

**Table 4: Statistical Analysis of Accessibility Scores of Users with Special Needs**

According to table 4, the probability of these results assuming the null hypothesis is 0.77. Therefore the null hypothesis is valid and the research hypothesis cannot be accepted. Thus, the website accessibility cannot be proved to be influenced by the individual impairment. However, the raw data depicted in table 1 reveals, one each individual users from visual / sight and mobility impairment groups disagree with the accessibility of sites. A site inaccessible for a user with sight disorder, due to the lack of readable fonts or alter text features, should be inaccessible for all users with the same level of sight disorder. Hence, the sites are inaccessible for users with visual and mobility impairments. Abanumy *et al* (2005) reports the same phenomenon among the e-government websites of Saudi Arabia and Oman. Therefore this could be generalized as a general phenomenon in website design. The following section formulates the accessibility guidelines for hotel industry website designers. Thus, these results should be subjected to further analysis to identify the exact design elements which prevent the accessibility of sites for users with visual and mobility impairments. Consequently, a web content analysis of five sample sites was conducted based on randomly selected criterions from W3C accessibility guidelines. The shortcomings identified with the web content have been used extending the findings of preliminary survey and discussed in the proceeding section.

## 6. Web Content Analysis

The content analysis was done using two different ways. Firstly, the overall accessibility of individual sites was evaluated using an open source web accessibility evaluation tool. **POWERMAPPER** has been selected for automated analysis. The reason for the selection of this particular tool was due to the flexibility of evaluation in terms of its interface as well as its ability to support web 2.0. The evaluation version is limited to checking 10 pages and images. The accessibility test was conducted against WCAG1, WCAG2 check points. Priority 1 checkpoints from W3C guidelines were abstracted (W3C Guidelines).Only, the priority 1 issues mentioned here, where – disabled users will find it impossible to use some pages. Four websites were randomly selected and an accessibility evaluation was undertaken.

### *6.1. Priority 1: Check Points*

- Provide equivalent alternatives to auditory and visual content.
- Don't rely on color alone.
- Use markup and style sheets and do so properly.
- Clarify natural language usage
- Create tables that transform gracefully.
- Ensure that pages featuring new technologies transform gracefully.
- Ensure user control of time-sensitive content changes.
- Ensure direct accessibility of embedded user interfaces.
- Design for device-independence.
- Use interim solutions.
- Use W3C technologies and guidelines.
- Provide context and orientation information.
- Provide clear navigation mechanisms.
- Ensure that documents are clear and simple.

(W3C Guidelines)

### *6.2. Accessibility Issues*

The overall accessibility ratings of five randomly selected websites of the hotels located in Sri Lanka are presented below. The names of web sites are not been used for ethical reasons.

| Sample | Issues (%) |
|---|---|
| Sample 1 | 72% |
| Sample 2 | 45% |
| Sample 3 | 63% |
| Sample 4 | 27% |
| Sample 5 | 18% |
| **Mean** | **45%** |

**Table 5: Automatic Accessibility Ratings of Websites**

The overall mean value of accessibility issues with the evaluated sites shows a significant rate (45%). More than average number of sites (3) shows a higher rate of accessibility issues. Although there are two samples records a very low accessibility rate, the other three samples shows a very significant rate ranging from 45 % to 72 %. Hence, it is notable that the sites have accessibility issues, and the detailed information on the specific categories of issues, is been analyzed to identify the specific shortcomings with regard to the web content. The different categories are been taken to see the different types of accessibility issues. The selection of categories has been made based on convenience to evaluation. Hence the categories given in the accessibility checking tool were adapted as it is.

| Category        | Issues (S1) | Issues (S2) | Issues (S3) | Issues (S4) | Issues (S5) |
|-----------------|-------------|-------------|-------------|-------------|-------------|
| Overall Quality | 2           | 3           | 6           | 5           | 7           |
| Errors          | 1           | 3           | 4           | 0           | 4           |
| Accessibility   | 1           | 3           | 4           | 5           | 7           |
| Compatibility   | 1           | 3           | 5           | 1           | 4           |
| Compliance      | 1           | 3           | 4           | 4           | 4           |
| Search          | 0           | 3           | 4           | 5           | 5           |
| Standards       | 1           | 3           | 4           | 5           | 7           |
| Usability       | 1           | 3           | 5           | 5           | 5           |

**Table 6: Details of Automatic Accessibility Ratings**

The detailed results show that the sites have significant issues with regard to the overall quality. Secondly, the sites do have considerable amount of errors, which greatly affects the accessibility of sites. Apart from this the significant issues are identified with the accessibility and usability of sites. Finally, the search results are another significant issue which prevents the sites to be accessible for users with different impairments.

## 6.3. Validation

W3C classify the accessibility guidelines under four main principles namely; (1) perceivable, (2) operable, (3) understandable, (4) Robust (W3C WAI, 2008)). One each guideline from the above four principles were randomly selected from the four main categories for analysis. The selected criterions are: (1) Text alternatives, (2) Navigable, (3) Readable, (4) Compatible. More than 60% of the sites reviewed do not have alt-text facility. This will have a significant effect for users with visual impairments. The navigation through pages is observed to be taking more time and this will definitely de-motivate the user. Especially, the location specification of the web pages has not been indicated during the course of navigating through pages. A very poor readability is observed in more than 60% of the sites analyzed. Especially, the sites do not have fonts which are visible to the users with sight impairments, which is mostly because of the inappropriate usage of color combination in the font selection. A specific test for all five samples was done to check the readability. Notably, readability is one of the significant issues which contribute towards the accessibility of sites for users with visual impairments. One hotel uses photo enlarging option to support visually impaired users.

Another hotel holds two options in their homepage. One is for customers with slow connections and other one is for speed connections. Basic product information and company information are presented in both sites. Although, the site compatible with slow connections have designed with less multimedia objects, which lacks to incorporate readable fonts in web design.

## 7. Implications

- *Adherence to principles:* The web content analysis showed a general weakness of websites with regard to lack of adherence to W3C accessibility guidelines. Especially, the alt-text facility, navigation and readability are the specific issues which do not meet the compliances of guidelines. Disregarding W3C WAI guidelines had led to poor design, which reduces the readability of websites to people with visual impairments. Also, the mismatch with W3C guidelines has prevented sites from being accessible to mobility users. This could be synthesized as a common weakness in most of the commercial websites from developing regions.

- *User-Centered Design:* There are arguments which claim that it is a danger to solely depending on the accessibility guidelines (Milne *et al*, 2005). Especially, that it would be difficult to cater for the needs of all known disabilities. On the other hand, the exact need of the user could be triggered solely by including users in the design process. There are a number of user centered design techniques that would go some way to reducing accessibility issues (Preece *et al*, Nielsen, 2004) This is a most common weakness in hotel websites. Therefore, it is recommended to adapt a user - centered approach in web design. Especially, it is vital to include users with visual and mobility impairments as well some elderly users in the design process, to ensure the needs diverse user groups are addressed. The development of design models and testing the design with end users would immensely increase the accessibility of websites. Although, the above best practices are well practiced by most of the web designers in the developed world, a major setback on accessibility concerns has been observed among the organizations located in the developing countries.

- *Re-ordering Search Results:* Re – ordering the search results according to user needs is an innovative method through which the issue related to of search could be addressed. The automated reordering could be implemented with the website search engines and this will complement a great deal of issues arising due to the lack of search optimization. Especially, implementing such a system would increase the accessibility of sites for users with differing needs and which will eventually lead to increased accessibility and consequently increased business outcomes. This in other way round will reduce the design exclusion and therefore this could lead to universal design paradigm where the needs attainment of customers would be very high and the consequent outcomes of the businesses in terms of economic and ethical basis would be relatively high. Therefore, search engine optimization is a viable implication for improving accessibility of e-commerce websites to users with differing impairments.

- *Portals with different speeds*: Although the hypothesis tests shows there is no co-relation between the accessibility and the type of internet connections, the results of this study has been obtained from a comparatively small sample for connection wise users groups. Especially, the sample for users with other connection types was comparatively small. On the other hand, there is a high probability for greater number of users from developing countries holding low speed connections. Hence, it is recommended to design websites with customized portals to match with the differential speed categories of users.

- ➢ *Accessibility Strategy*: A lack of accessibility strategy among the hoteliers has been clearly scrutinized through the above research. Especially, the websites are not having any proper standard with regards to the accessibility for users with differing needs. This could be scrutinized through improper color combination and lack of clarity of fonts. Particularly, the sites are designed on the convenience of designers than of users. The lack of mutual contribution between academia and industry and poor awareness level on accessibility issues could be the main reasons for the above phenomenon in developing countries. Christian (2009) reported a similar phenomenon with Australian small hotels and suggested to adapt a web strategy to avoid lapse. However, an accessibility strategy would be a sensible solution for the above issue in the context of this research. As part of accessibility strategy, the organizations could organize knowledge transfer sessions for web designers and walkthroughs with users. This will help the designers to grasp a high level understanding of the accessibility issues as well as to elicit the exact user requirements with regard to accessibility.

## 8. Evaluation & Conclusions

The study investigated on implications for improving the accessibility of e-commerce websites in developing countries. The study compiled with a case study of Sri Lankan hotel websites. The results indicate poor accessibility for users with visual and mobility impairments. Poor readability and less navigable pages in the websites were scrutinized as the principle reasons which prevent accessibility of sites for users with visual and mobility impairments. Also, the sites have been observed not adhering to W3C accessibility principles. The implications includes adherence to principles, user centered design, developing design frameworks, portals with different speeds and an accessibility strategy.

The study lays a foundation for web accessibility for websites research in developing countries. Especially, this is an explorative study for Sri Lankan hotel industry and the results will shed light on future research avenues towards developing specific design models for accessible web design. Also, the outcome could be extended further by developing testable prototypes to address differencing user needs, testing which with different user groups would reveal specific contextual needs of tourists with special needs.

On the other hand, this study could be criticized for a number of limitations. Firstly, the research hypothesis was developed solely based on a predisposition that the economic diversity of the sample could be exactly maintained through including users with different connection types. It could be argued that there could be other different parameters which could form a clear foundation for accessibility research of users from different economic backgrounds. Secondly, the selection of sample consists of generic users instead of tourists. However, it is defendable as the accessibility is same for a tourist and non-tourist. In addition to this, the non equal distribution of samples among the sample groups could be argued as another inherent weakness of this research.

In conclusion this research is just a beginning for Sri Lankan hotel industry. Further research needs to be undertaken to extend the findings on factors influencing the accessibility of sites for users from different economic, geographic and physical backgrounds. However, this particular study remains as beginning for Sri Lankan hotel industry, while adding some facts to the existing knowledge pool of accessibility research literature for the context of developing countries. The inaccessibility of sites in developing regions will immensely affect the economic development of the region through preventing a considerable amount of foreign exchange which earned through e-commerce business. Hence, the implications developed as part of this study will serve as a positive step towards the enhanced economy of developing world.

**Appendices**

| Type of Connection | Users |
|---|---|
| Dial Up | 3 |
| ADSL | 49 |
| Broadband | 54 |
| Any Other | 1 |

**Table 7: Connection wise Responses**

| Special Needs | Users |
|---|---|
| Yes | 17 |
| No | 87 |

**Table 8: Users with Special Needs**

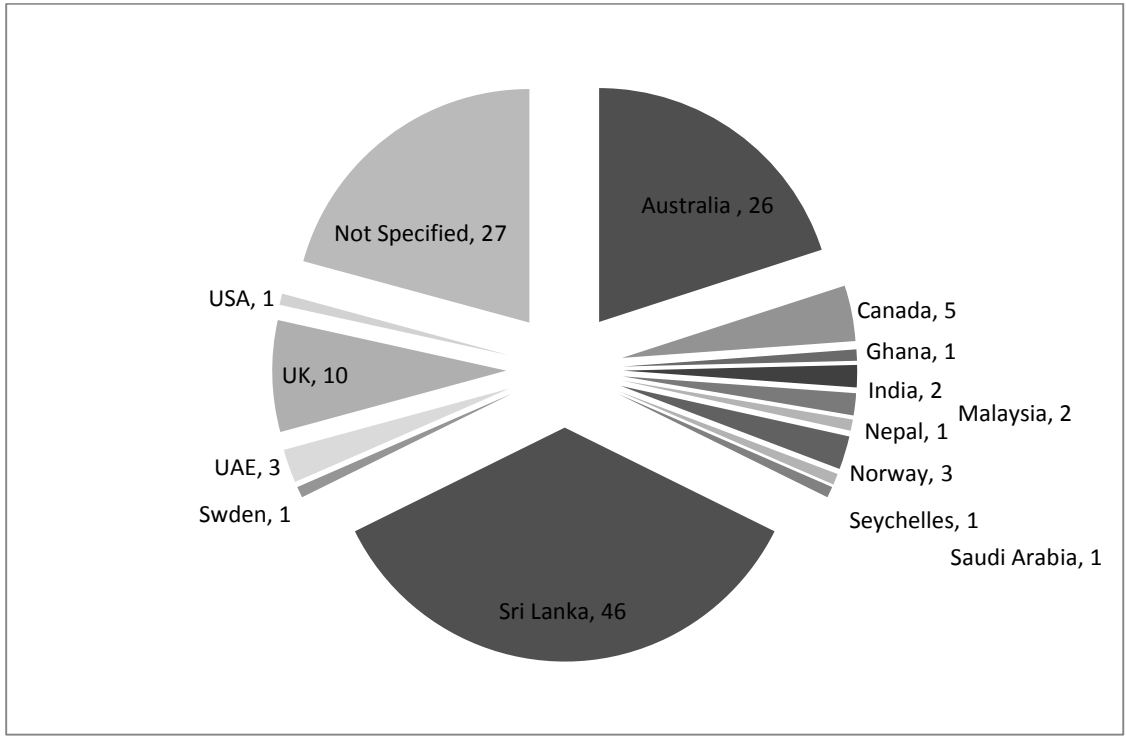

**Figure 1:** Country wise Respondents Detail